\documentclass[preprint,aps]{revtex4}
\usepackage{graphicx}
\usepackage{bm}
\usepackage{fancyhdr}
\usepackage{amsmath}
\usepackage{amsfonts}
\usepackage{amsbsy}
\usepackage{amssymb}
\usepackage[mathscr]{eucal}
\usepackage{color}
\usepackage{soul}
\usepackage{mathtools}
\graphicspath{{./Figures/}}

\begin{document}

\title{Nonequilibrium statistical mechanics of money/energy exchange models}

\author{Maggie Miao}
\affiliation{Department of Chemistry and Oden
Institute for Computational Engineering and Sciences, The University of Texas at Austin, Austin, Texas 78712, United States}

\author{Kristian Blom}
\affiliation{Mathematical bioPhysics group, Max Planck Institute for Multidisciplinary Sciences, G\"{o}ttingen 37077, Germany}

\author{Dmitrii E. Makarov}
\email{makarov@cm.utexas.edu}
\affiliation{Department of Chemistry and Oden
Institute for Computational Engineering and Sciences, The University of Texas at Austin, Austin, Texas 78712, United States}
\date{\today}

\date{\today }
\begin{abstract}
Many-body dynamical models in which Boltzmann statistics can be derived directly from the underlying dynamical laws without invoking the fundamental postulates of statistical mechanics are scarce. Interestingly, one such model is found in econophysics and in chemistry classrooms: the money game, in which players exchange money randomly in a process that resembles elastic intermolecular collisions in a gas, giving rise to the Boltzmann distribution of money owned by each player. Although this model offers a pedagogical example that demonstrates the origins of Boltzmann statistics, such demonstrations usually rely on computer simulations -- a proof of the exponential steady-state distribution in this model has only become available in recent years. Here, we study this random money/energy exchange model, and its extensions, using a simple mean-field-type approach that examines the properties of the  one-dimensional random walk performed by one of its participants. We give a simple derivation of the Boltzmann steady-state distribution in this model. Breaking the time-reversal symmetry of the game by modifying its rules results in non-Boltzmann steady-state statistics. In particular, introducing “unfair” exchange rules in which a poorer player is more likely to give money to a richer player than to receive money from that richer player, results in an analytically provable Pareto-type power-law distribution of the money in the limit where the number of players is infinite, with a finite fraction of players in the ``ground state'' (i.e., with zero money). For a finite number of players, however, the game may give rise to a bimodal distribution of money and to bistable dynamics, in which a participant's wealth jumps between poor and rich states. The latter corresponds to a scenario where the player accumulates nearly all the available money in the game. The time evolution of a player's wealth in this case can be thought of as a ``chemical reaction'', where a transition between ``reactants'' (rich state) and ``products'' (poor state) involves crossing a large free energy barrier. We thus analyze the trajectories generated from the game using ideas from the theory of transition paths, and highlight non-Markovian effects in the barrier crossing dynamics.   
\end{abstract}
\maketitle

\section{Introduction}\label{SecI}
Simple asset exchange games have been used as models of wealth distribution in econophysics \cite{angle1986surplus,greenberg2023twentyfive, RevModPhys.81.1703, dragulescu2000statistical, 10.1119/1.4807852, Banerjee_2010, bennati1988metodo, bennati1993metodo, ispolatov1998wealth, PhysRevE.104.014151, PhysRevE.104.014150,DRAGULESCU2001213, A_Silva_2005, BANERJEE200654}.
 Such models are also of interest as relatively simple many-body systems whose time evolution can be shown, explicitly and without resorting to the fundamental postulates of statistical mechanics, to give rise to the Boltzmann distribution \cite{scalas2006statistical, lanchier2017rigorous, lanchier2018rigorous, lanchier2019rigorous, lanchier2022distribution}. Indeed, microscopic models where fundamental laws of statistical mechanics can be proven without invoking statistical assumptions are scarce, and such proofs are often complex -- see, e.g.,  \cite{billiardsReview, simanyi2003proof, simanyi2009conditional}. 
 
In contrast, the onset of the Boltzmann distribution in the simple game where $N\gg1$ players exchange a fixed amount of money in a random direction can be understood using simple arguments. In this game, which is known in econophysics as the Bennati-Dregulescu-Yakovenko game \cite{dragulescu2000statistical, RevModPhys.81.1703}, and which is also used by physical chemistry teachers to illustrate various concepts of statistical mechanics \cite{michalek2006give}, a randomly selected pair of players exchange a single money unit in a random direction. A physical counterpart of this game corresponds to $N$ ``molecules" with (quantum) harmonic oscillator energy spectra, and with pairs of oscillators exchanging one quantum of energy at random such that the total energy is conserved. The ``fair" version of this game corresponds to time-reversible exchange dynamics where the random exchange direction is unbiased, leading to a Boltzmann distribution of money/energy \cite{dragulescu2000statistical, scalas2006statistical}. In contrast, ``unfair'' versions of the game with a bias in the exchange direction give rise to interesting scenarios with non-Boltzmann money statistics \cite{FeiCao2023, scafetta2002paretos, SCAFETTA2004338}.

Here, we present analytical results for a class of exchange games, where ``unfair'' exchange rules correspond to broken time-reversal symmetry. We start with an elementary derivation of the Boltzmann distribution as the steady-state result of the fair exchange game with $N\rightarrow \infty$ players. We note that another derivation of this distribution was recently provided in \cite{lanchier2017rigorous, lanchier2018rigorous, lanchier2019rigorous}. We then introduce ``unfair'' versions of the game, where money exchange, while still probabilistic, is biased to increase or decrease the wealth of the richer or poorer players (we call them rich-biased and poor-biased games, as in \cite{FeiCao2023}). A key feature of these games is that they violate time-reversal symmetry, leading to nonequilibrium dynamics with dissipative cycles. As a result, the steady-state probability $p_m$ of the money $m$ belonging to a player no longer has an exponential dependence on $m$ predicted by the Boltzmann law. 

The $m$-dependence of $p_m$ is qualitatively different depending on whether the game is poor-biased or rich-biased. In the former case, we find a bell-shaped distribution $p_m$ centered around the mean. The latter case is more interesting. In the limit where both the number of players $N$ and the average amount of money per player $\langle m \rangle$ approach infinity, the function $p_m$ is a power law, while in the limit $N\rightarrow\infty$ with $\langle m \rangle$ fixed, the distribution $p_m$ follows a power law at intermediate values of $m$ (where this distribution turns out to be independent of the value of $\langle m \rangle$) and becomes exponential in the limit $m\rightarrow\infty$. Another interesting feature of this regime is that the fraction of players with zero money (or, equivalently, the fraction of particles in the ground state) always remains finite. For finite $N$ (in the rich-biased case) we find that finite-size effects dominate the dynamics, resulting in a bimodal wealth distribution, with the money belonging to each player undergoing non-Markovian bistable dynamics switching between poor and rich states, and with the rich state occurring when the player accumulates nearly the entire money in the system.

This paper is organized as follows: In Section~\ref{SecII} we describe details of the model. Section~\ref{SecIII} discusses the ``fair game'' case and gives simple arguments explaining the Boltzmann distribution of money in this case. Unfair exchange games are introduced in Section~\ref{SecIV}, where the connection between unfairness and time-reversal symmetry breaking is shown. The general mean-field solution to such games is introduced in Section~\ref{SecV}. Section~\ref{SecVI} reports on analytical results for the case of a rich-biased game in the limit of infinite number of players, while Sections~\ref{SecVII} - \ref{SecIX} discuss finite-size effects in this game using both simulations and analytic theory. Section~\ref{SecX} discusses the timescales to reach the steady state, and Section~\ref{SecXI} concludes by highlighting the most important findings of this work. Supplementary Material provides further details of the analytical theory used to predict the wealth distribution in the case of a finite number of players.


\section{Model}\label{SecII}

The model studied here is described, more precisely, as follows: Each of the $N$ players (molecules) has $m_i$ money (or energy) units, where the index $i=1,\dots, N$ enumerates the players. The total amount of money, 
\begin{equation}
M=\sum_{i=1}^N m_i
\end{equation}
is conserved and fixed by the initial conditions. At every step, a pair, say $i$ and $j$, is selected at random; If $m_i>0$ and $m_j>0$, then these players/molecules exchange money/energy,  \begin{equation} \label{exchange}
m_i\rightarrow m_i \pm 1, m_j \rightarrow m_j\mp 1
\end{equation}
with the signs ``+" and ``-" determining which player receives the money selected according to certain probabilistic rules to be specified below. When only one direction of exchange is possible, the exchange becomes deterministic. For example, if $m_i=0$ and $m_j \ne 0$, the result of the exchange is $m_i \rightarrow 1$, $m_j \rightarrow m_j-1$. If $m_i=0$ and $m_j=0$, then no exchange takes place, $m_i \rightarrow m_i$, $m_j \rightarrow m_j$.
We note that these rules may be viewed as unrealistic from an economics perspective: for example, it could be more realistic to assume that a player reaching the zero-money state leaves the game \cite{ispolatov1998wealth}. But since we are more interested in molecular consequences than economic implications of the model here, the assumption of conservation of the number of particles/players $N$ is more natural. 

We are interested in the steady-state probability $p_m$ that a player has $m$ money units, particularly in the limit, 
\begin{equation}
N\rightarrow \infty, \ \frac{M}{N}=\langle m \rangle,
\end{equation}
with the average amount of money per player $\langle m \rangle$ being a given parameter specified by the initial conditions. 

\section{Fair exchange leads to Boltzmann distribution}\label{SecIII}

In the fair game, when both directions of exchange (i.e. both signs in Eq.~\eqref{exchange}) are possible, they are chosen with equal probabilities. Note that the dynamics under the fair game rules is  time-reversible - see Section \ref{SecIV}. 

\begin{figure}
    \centering
    \includegraphics[width=0.7\textwidth]{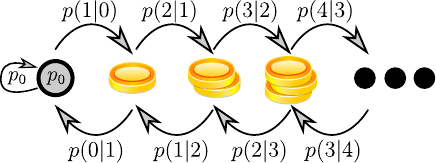}
    \caption{Kinetic scheme for a single player in the Bennati-Dregulescu-Yakovenko game. }
    \label{fig:scheme}
\end{figure}

We now give a simple physical explanation why fair exchange rules lead to the Boltzmann distribution in the steady-state. Focus on a single participant of the game, and consider the evolution of the amount of money $m$ owned by this player. This quantity undergoes a one-dimensional random walk, as shown in Fig.~\ref{fig:scheme}.  Let $p(m+1|m)$ and $p(m-1|m)$ be the conditional probability that a player with amount $m$ will increase/decrease its amount by $1$ (see Fig.~\ref{fig:scheme}) such that the walker steps right/left. Given the fairness of the game, $m$ is equally likely to increase/decrease upon encountering another player with a nonzero amount of money. When encountering a player with zero amount of money, however, our participant cannot receive money and must lose money instead (assuming $m \geq 1$). The probability that the player encountered is broke is $p_0$, and the probability that this player is not $1-p_0$; therefore we have
\begin{equation}
p(m-1|m)=\frac{1}{2} (1-p_0) +p_0 = \frac{1+p_0}{2}, \ {\rm for} \ m\geq 1.
\label{pminfair}
\end{equation}
Similarly, we find
\begin{equation}
p(m+1|m)=\frac{1-p_0}{2}, \ {\rm for} \ m\geq 1.
\label{pplusfair}
\end{equation}
For $m=0$ we obviously have $p(-1|0)=0$, as $m$ cannot become negative. To find $p(1|0)$ we note that the walker takes a step to the right ($m$ increases) provided that the player encountered is not broke (with probability $1-p_0$) or remains unchanged otherwise (with probability $p_0$). Thus, $p(1|0)=1-p_0$. Finally, if a player with no money encounters another player that is broke, the probability of such an encounter is $p_0 \equiv p(0|0)$ (Fig.~\ref{fig:scheme}). 

The steady-state probabilities $p_m$ must satisfy the detailed balance condition, i.e.,
\begin{equation}
p_m p(m\pm 1|m)=p_{m\pm 1}p(m|m\pm 1),  
 \label{detailed balance}
\end{equation}
from which we find:
\begin{equation} 
p_m = 2 p_0 \left( \frac{1-p_0}{1+p_0}\right)^{m}, \ {\rm for} \ m\geq 1.
 \label{Boltzmann}
\end{equation}
This is the Boltzmann distribution, with $p_m$ exponentially decaying with increasing $m$. Importantly, the probabilities satisfying Eq.~\eqref{Boltzmann} automatically satisfy the normalization condition
\begin{equation}
\sum_{m=0}^{\infty} p_m  = 1.  
\end{equation}
The value of $p_0$, then, must be determined by the initial amount of money in the system
\begin{equation}
\sum_{m=0}^{\infty} m p_m = \langle m \rangle.
\label{average constraint}
\end{equation}
In combination with Eq.~\eqref{Boltzmann}, this gives
\begin{equation}
p_0=\sqrt{\langle m \rangle^2 +1}-\langle m \rangle,
\end{equation}
or
\begin{equation}
\langle m \rangle=\frac{1}{2p_0}-\frac{p_0}{2}.
\label{money vs p0}
\end{equation}
The occupancy of the zero-money state (i.e. of the single-molecule ``ground state") $p_0$ is thus a monotonically decreasing function of the average money per player. In the limit $\langle m \rangle\gg 1$, Eq.~\eqref{money vs p0} gives $p_0\approx 1/2\langle m \rangle$, and Eq.~\eqref{Boltzmann} can be written, approximately, as
\begin{equation}
p_m \approx \beta \exp{(-\beta m)},
\label{Boltzmann1}
\end{equation}
with 
\begin{equation} \label{Boltzmann2}
1/\beta = \langle m \rangle.     
\end{equation}
As expected, the ``temperature" of the Boltzmann distribution is, in this limit, equal to the average money/energy per player/molecule.

In the opposite limit of "scarce resources", $\langle m \rangle \ll 1$, we find from Eq.~\eqref{money vs p0}, 

\begin{equation}
    p_0\approx 1- \langle m \rangle
\end{equation}
and, then, from Eq.~\eqref{Boltzmann},
\begin{equation}
p_m\approx 2^{1-m} \langle m \rangle^m, \  m\ge 1.
\end{equation}

We note that, for a finite number of players $N$, Eq.~\eqref{Boltzmann} is not exact: Indeed, in contradiction to this equation, the probability $p_m$ must be equal to zero for $m>M$. This case can be analyzed, systematically, using a local equilibrium approximation described in Section \ref{SecVIII} and in the Supplementary Material. In practice, however, assuming $\langle m \rangle \ll M$, the probability $p_m$ predicted by Eq.~\eqref{Boltzmann} is vanishingly small for any $m$ such that $m\gg \langle m \rangle$. In other words, our theory works under the assumption that each player can only amass a vanishingly small (comparable to $\langle m \rangle$) fraction of the total money pool $M=N\langle m \rangle$, which is true for $N\gg1$. As will be seen below, however, this assumption is not necessarily satisfied when the rules of the game are changed to be unfair and when $N$ is finite.

Finally, let us point out another important feature of Eqs.~\eqref{Boltzmann1}-\eqref{Boltzmann2}  (but not of Eq. \eqref{Boltzmann}): The distribution of money is independent of certain details of the game. More precisely, recall that, the ``unit'' of money in our game is the same as the amount of money exchanged in each encounter between players. Let $\Delta \mu$ be the actual exchanged money, say, in dollars or cents, and let $\mu = m\times \Delta \mu$ be the money owned by a player. Then it follows from Eqs.~\eqref{Boltzmann1}-\eqref{Boltzmann2} that the \emph{probability density} of the wealth $\mu$, measured at a sufficiently low resolution such that the discreteness of $\mu$ is irrelevant, is given by the exponential law:
\begin{equation}
p_{\mu}(\mu) \approx p_m \frac{dm}{d\mu}\bigg|_{m=\mu/\Delta\mu} =\frac{1}{\langle \mu \rangle}\exp{(-\frac{\mu}{\langle \mu \rangle})},
\label{Boltzmann3}
\end{equation}
which is independent of the money exchanged ($\Delta \mu$). In other words, regardless of whether players exchange dollars or cents, their wealth distribution will be the same as long as the average wealth per player is the same. In general, this is not the case (e.g. for the more general result of Eq.~\eqref{Boltzmann}).

\section{Unfair exchange implies broken time reversal symmetry}\label{SecIV}
We now introduce a class of ``unfair" exchange models, in which the direction of exchange between two players with amounts of money $m$ and $m'$ depends on $m$ and $m'$. Specifically, let $\phi_+(m,m') \equiv p(m+1,m'-1|m,m')$ be the (conditional) probability that a player with $m$ money units will accept money upon encountering a player with $m'$ money units. Assuming $0<m,m'<M$, the probability that this player will give money to the other player is, then, $\phi_-(m,m')\equiv p(m-1,m'+1|m,m')=1-\phi_+(m,m')$, and, by symmetry, we also have $\phi_\pm(m,m')=\phi_\mp(m',m)$.
In what follows, we will focus on a particular example of such a model, where the exchange probabilities are determined by the sign of the difference $m-m'$. Specifically:
\begin{equation} \label{unfair models}
\phi_+(m,m')= 
\begin{cases}
\phi, & \quad 0<m'<m<M,\\
1-\phi, & \quad 0<m<m'<M,\\
1/2, & \quad 0<m=m'<M, \\
0, & \quad m=M,\\
1, & \quad 0=m<m',\\
0, & \quad m=m'=0.
\end{cases}
\end{equation}
The quantity $\phi$ is the probability that a wealthier player encountering a poorer one (but still with nonzero money) will receive money from the latter. Thus, it is a measure of unfairness of the game, with $\phi=1/2$ corresponding to the fair game discussed in the preceding Section. When $\phi<1/2$, the game will tend to equalize the wealth of the players. When $\phi>1/2$, it will increase inequality. Intuitively, time reversible exchange dynamics should correspond to the fair game case: for example, if we replace players with colliding molecules and money with their energies, then a trajectory of two colliding trajectories and its time-reverse result in the same amount of energy being exchanged in the opposite directions.    

\begin{figure}[h!]
    \centering
    \includegraphics[scale=.7]{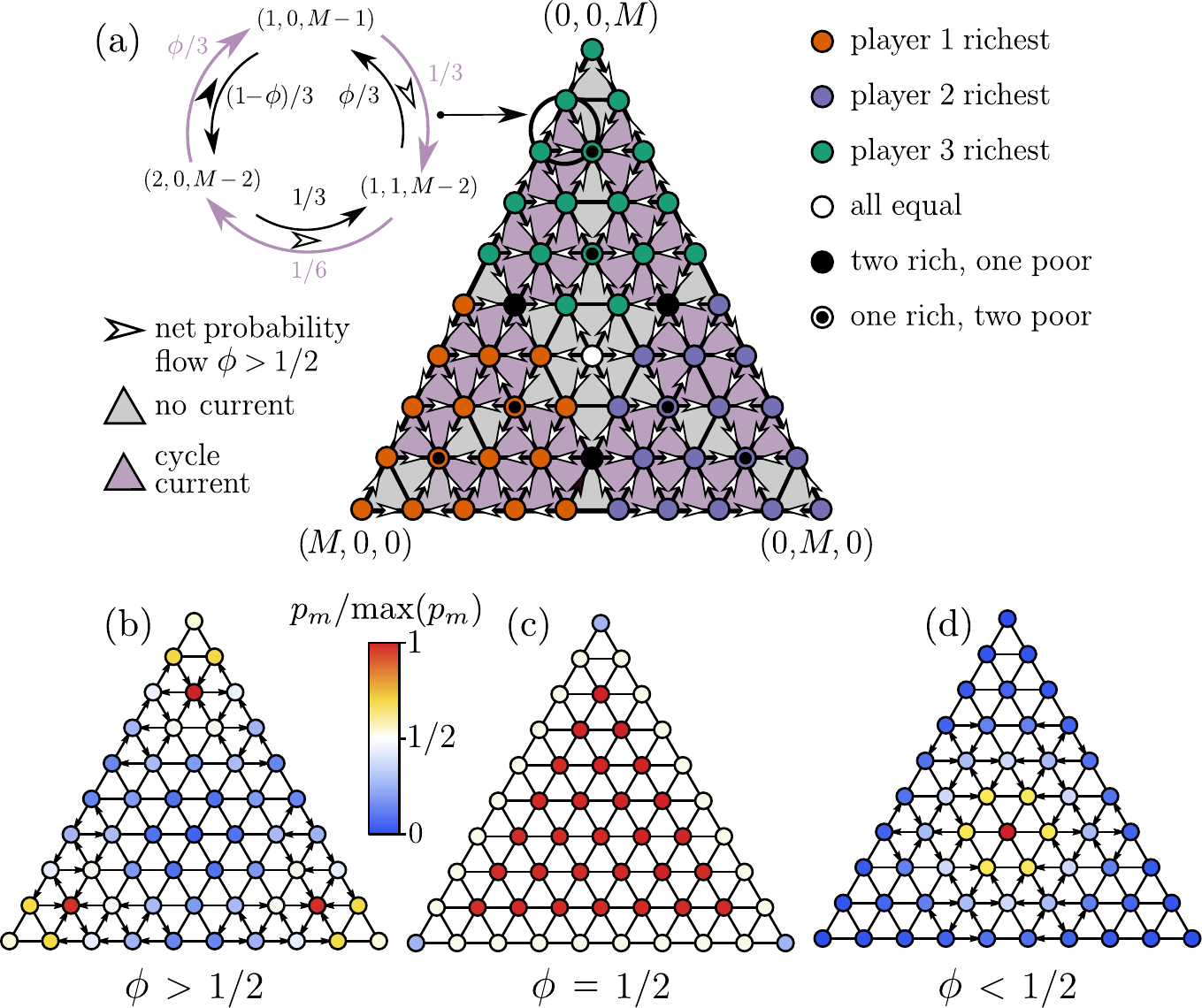}
    \caption{(a) Full kinetic scheme of the money game with $N=3$ players and $M=9$ units of money ($\langle m\rangle=3$). In the upper left corner, we show a cycle which contains a net probability flow in the clockwise direction for $\phi>1/2$. For $\phi<1/2$ the cycle current is reversed, while for $\phi=1/2$ there are no detailed-balance-violating cycles. (b-d) Steady-state probabilities for $\phi=0.3$ (b), $\phi=1/2$ (c), and $\phi=0.7$ (d), obtained from simulations. Black arrows indicate steady-state probability fluxes. It can be shown that, for $\phi = 1/2$, most states in this kinetic scheme have the same equilibrium population (this follows from detailed balance conditions for different pairs of states). This, however, is not the case for $\phi \ne 1/2$ where detailed balance does not hold.}
    \label{fig:full kinetics}
\end{figure}

To quantify the above statement more precisely, note that the time evolution of the state vector $(m_1,m_2,m_3...,m_N)$ of the system is described by a discrete-time master equation, with transition probabilities specified by Eq.~\eqref{unfair models} multiplied by the probability that two specific players meet. As an example, a kinetic scheme describing this system is shown in Fig.~\ref{fig:full kinetics} for the case $N=3$. Importantly, the dynamics satisfies detailed balance and is time-reversible only for the  fair game with $\phi=1/2$. Indeed, according to Kolmogorov's cycle criterion \cite{kelly2011reversibility, kolmogoroff1936theorie}, detailed balance is violated if there is a cyclic sequence of microscopic states such that the product of the transition probabilities in the clockwise direction is different from that taken in the counterclockwise direction. Examples of such cycles, for $\phi \ne 1/2$, are highlighted in Fig.~\ref{fig:full kinetics}. And since any $3$-player game can be embedded in a game with $N\geq3$ players, the example in  Fig.~\ref{fig:full kinetics} suffices to prove that detailed-balance is broken when $\phi\neq1/2$.

Various other "unfair" exchange models have been studied in the literature \cite{FeiCao2023, scafetta2002paretos, SCAFETTA2004338}. In \cite{FeiCao2023}, the exchange direction is always set from a ``giver" to a ``receiver", and the bias is introduced in the probability of selecting a ``giver" and a ``receiver", which depends on the amount of money each person has (see Fig.~1 in \cite{FeiCao2023}). In \cite{scafetta2002paretos, SCAFETTA2004338} the bias in the exchange direction does not only depend on the sign difference of $m-m^{'}$ (as we have here), but also on the magnitude of the difference (see i.e., Eq.~(16) in \cite{SCAFETTA2004338}). The latter, however, reduces the analytical tractably of the model. 

\section{Mean-field theory: evolution of wealth as a Markovian random walk}\label{SecV}
Consider the one-dimensional random walk shown in Fig.~\ref{fig:scheme}. Within mean-field theory, the transition probabilities of stepping right and left are computed as weighted averages of the microscopic transition probabilities,
\begin{equation}
p(m\pm 1|m)=\sum_{m'}\phi_\pm(m,m')p_{m'}.  
\end{equation}
For the model specified by Eq.~\eqref{unfair models}, then, we can write
\begin{equation} \label{pPlus}
p(m+1|m)=\phi \sum_{m'=1}^{m-1}p_{m'} + p_{m}/2 + (1-\phi)\sum_{m'=m+1}^{\infty} p_{m'}.
\end{equation}
In writing Eq.~\eqref{pPlus} it was assumed that the probabilities $p_m$ decay to zero, as $m\rightarrow \infty$, fast enough that the upper summation limit can be extended to infinity; this assumption allows us to disregard large values of $m$ where, for example, the money $m'$ owned by another player cannot be greater than $m$ if $m>M/2$. 

Introducing now 
\begin{equation}
\rho_m=\sum_{m'=0}^{m-1}p_{m'},
\end{equation}
we can rewrite Eq.~\eqref{pPlus} as:
\begin{equation} \label{pPlusNew}
p(m+1|m)=(2\phi - 1) \rho_m + 1-\phi+(\phi-1/2)p_m -\phi p_{0}.
\end{equation}
Similarly, we have 
\begin{equation} \label{pMinusNew}
p(m-1|m)=1-p(m+1|m)=-(2\phi - 1) \rho_m + \phi-(\phi-1/2)p_m +\phi p_{0}.
\end{equation}
Using detailed balance, given by Eq.~\eqref{detailed balance}, the steady-state probabilities $p_m$ can now be determined, starting from $p_0$,  by iterating the following map:
\begin{align}
    p_{m+1}&=p_m \frac{(2\phi - 1) \rho_m + 1-\phi+(\phi-1/2)p_m -\phi p_{0}}{-(2\phi - 1) (\rho_m+p_m) + \phi-(\phi-1/2)p_{m+1} +\phi p_{0}}, \label{map1} \\
    \rho_{m+1}&=\rho_m+p_m, \label{map2}
\end{align}
which holds for $m\ge 1$. Note that Eq.~\eqref{map1} contains $p_{m+1}$ both on the RHS and LHS - each iteration, then, involves solving a quadratic equation for $p_{m+1}$. 

Eqs.~\eqref{map1} and \eqref{map2} are not valid for $m=0$;  writing the detailed balance condition between states with $m=0$ and $m=1$, we obtain the first step of the map (cf. Fig.~\ref{fig:scheme}):
\begin{equation}
    p_0(1-p_0)=p_1p(0|1)=p_1 [-(2\phi - 1) \rho_1 + \phi-(\phi-1/2)p_1 +\phi p_{0}],
\end{equation}
or, using $\rho_1=p_0$,
\begin{equation} \label{map3}
    p_1=\frac{p_0(1-p_0)}{-(2\phi - 1) p_0 + \phi-(\phi-1/2)p_1 +\phi p_{0}}.
\end{equation}
Again, Eq.~\eqref{map3} results in a quadratic equation that allows one to determine $p_1$ starting from $p_0$. 

It is worth noting that the use of the detailed balance condition, given by Eq.~\eqref{detailed balance}, is justified \emph{even} when the system, viewed microscopically, is in a nonequilibrium steady state rather than in equilibrium, as is the case for $\phi \ne 1/2$ (see the previous Section). Indeed, the {\em linear} kinetic scheme of Fig.~\ref{fig:scheme} always obeys detailed balance. This scheme describes a projection of nonequilibrium dynamics in a $N$-dimensional space onto a single degree of freedom $m$, where - as is often the case for projected dynamics (see, e.g., \cite{AljazFRET,Carlos,KineticHysteresis,Roldan,Seifert}) - the nonequilibrium character of the underlying process is hidden, although it affects the distribution $p_m$.       

\begin{figure}
    \centering
    \includegraphics[scale=0.55]{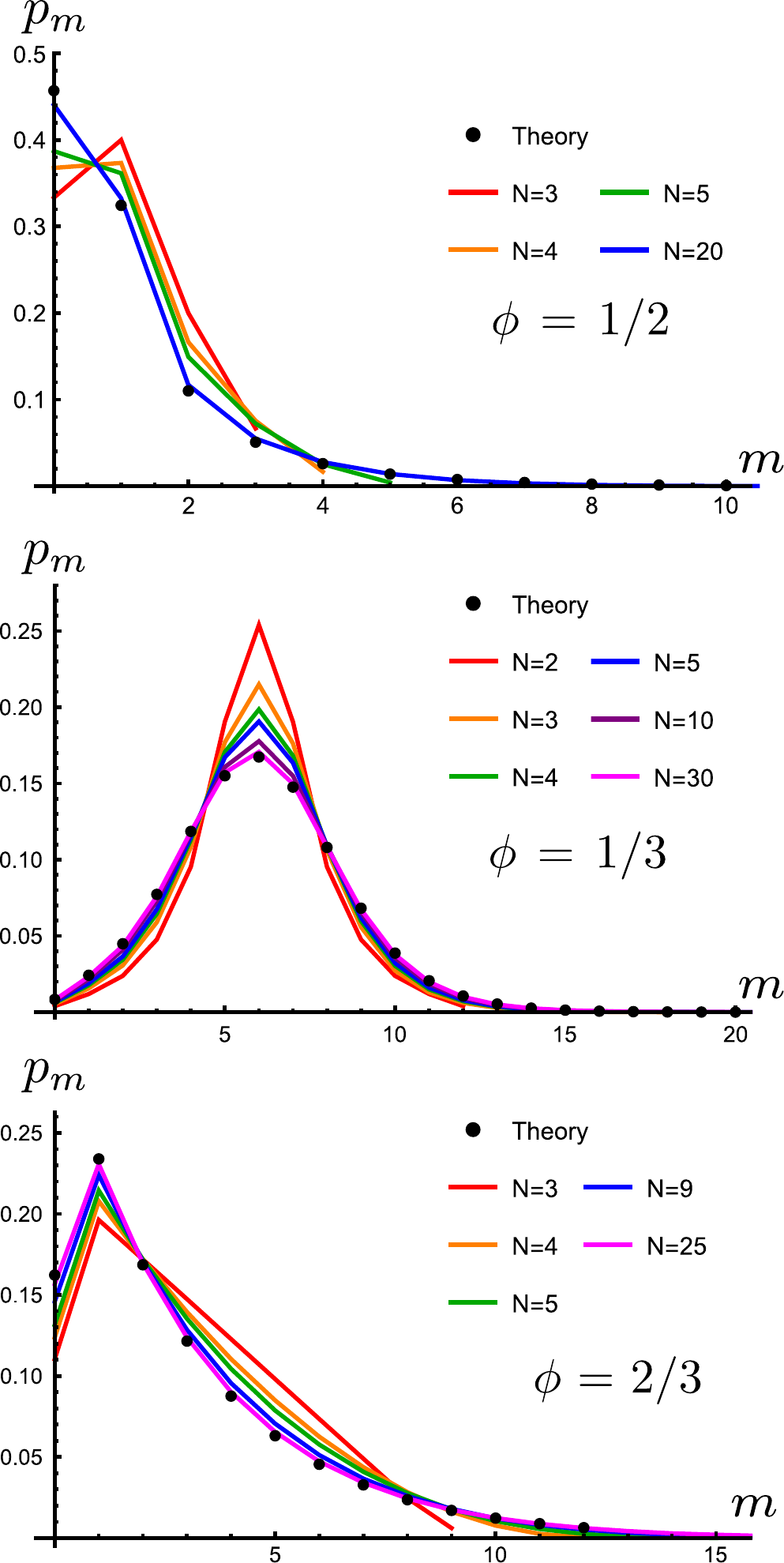}
    \caption{A comparison of Eqs.~\eqref{map1}-\eqref{map3} with simulations. The probability distributions for, from top to bottom, a $\phi=1/2$, $\langle m\rangle=3$ game, a $\phi=1/3$, $\langle m\rangle=6$ game, and a $\phi=2/3$, $\langle m\rangle=1$ game. The predictions of Eqs.~\eqref{map1}-\eqref{map3} are shown as black dots, while the simulation results are represented by continuous lines.}
    \label{fig:mean field}
\end{figure}

Figure \ref{fig:mean field} illustrates the performance of our theory in comparison with Monte Carlo simulations of money exchange. For $\phi=1/2$, the theory is identical to that of Section \ref{SecIII} leading to a probability distribution described by Eq.~\eqref{Boltzmann}. For $\phi<1/2$, the money exchange is more likely to proceed in the direction from a wealthier player to the poorer one; as a result, the  distribution $p_m$ has a peak centered around the average amount $\langle m\rangle$, which was also observed in \cite{scafetta2002paretos, SCAFETTA2004338}. This qualitative property of the distribution is captured by the theory, which becomes increasingly more accurate as $N$ increases. For $\phi>1/2$, the situation is more complicated. As will be seen in Section \ref{SecVII}, if $\langle m\rangle$ is large enough, then the distribution $p_m$ becomes bimodal, a qualitative feature not captured by our theory. The parameters in Fig.~\ref{fig:mean field} are chosen such that this is not the case. See Section \ref{SecVII} for a further discussion.  

\section{The $\phi>1/2$, $N=\infty$ case}\label{SecVI}

The most interesting regime is $\phi > 1/2$, where money transfer from poorer players to wealthier ones is favored. We start with considering the behavior of $p_m$ satisfying Eq.~\eqref{map1} in the limit $m\rightarrow \infty$. Since the distribution $p_m$ is normalized, we must have $$\lim_{m\rightarrow \infty}p_m = 0.$$ Moreover, we have $$\lim_{m\rightarrow \infty}\rho_m = \sum_{m=0}^\infty p_m=1.$$ Inserting the above two relations into Eq.~\eqref{map1}, we see that the ratio $p_{m+1}/p_m$ approaches a constant for $m\rightarrow\infty$, 
\begin{equation} \label{pRatio}
    \lim_{m\rightarrow \infty}\frac{p_{m+1}}{p_m}=\frac{\phi(1-p_0)}{1-\phi(1-p_0)}.
\end{equation}
On the other hand, $p_m$ can only vanish in the limit $m\rightarrow\infty$ if $p_{m+1}/p_m<1$. It then follows from Eq.~\eqref{pRatio} that a normalized steady-state solution exists only if 
\begin{equation} \label{normCond}
    p_0\ge p_c \equiv 1-\frac{1}{2\phi}.
\end{equation}
Hence, for $\phi>1/2$ we find $p_{c}>0$. In other words, in the rich-biased game, the fraction of players with zero money remains finite regardless of the total pool of money in the game. 

What happens when $p_0$ approaches the minimum possible value $p_c$ from above? To answer this question, let us write the map of Eq.~\eqref{map1} in the following form, 
\begin{equation} \label{map epsilon}
p_{m+1}=p_m \frac{2\eta (\tilde{\rho}_m-1)+1-\epsilon(\eta+1)}{-2\eta (\tilde{\rho}_{m+1}-1)+1+\epsilon(\eta+1)},     
\end{equation}
where
$$\eta=2\phi-1,$$
$$\tilde{\rho}_m=\rho_m + \frac{p_m}{2},$$ and $$\epsilon = p_0-p_c$$
For $m\rightarrow\infty$, we have $1-\tilde{\rho}_m\rightarrow0$, and thus in order to study the tails of the distribution $p_m$ we have to consider two small parameters, $\epsilon$, and $1-\tilde{\rho}_m$. Depending on the relationship between the two, there are two cases: \\{\bf Case A:\hspace{0.1in} $\epsilon \gg 1-\tilde{\rho}_m$}. Eq.~\eqref{map epsilon} can then be approximated by   
\begin{equation} \label{exp tail}
 p_{m+1}\approx p_m \frac{1-\epsilon(\eta+1)}{1+\epsilon(\eta+1)},     
\end{equation}
leading to an exponential tail for the distribution $p_m$, 
\begin{equation} \label{exp tail1}
    p_m\propto \left[ \frac{1-\epsilon(\eta+1)}{1 +\epsilon(\eta+1)} \right]^m.
\end{equation}
\\{\bf Case B:\hspace{0.1in} $\epsilon \ll 1-\tilde{\rho}_m$}. Eq.~\eqref{map epsilon} can now be rewritten as  
\begin{equation} \label{zero eps}
p_{m+1}\approx p_m \frac{2\eta (\tilde{\rho}_m-1)+1}{-2\eta (\tilde{\rho}_{m+1}-1)+1}\approx p_{m}(1-4\eta (1-\tilde{\rho}_m)).     
\end{equation}
Treating $m$ as a continuous variable, we can further approximate the above equation by  
$$
p_{m+1}/p_m \approx 1+p_m^{-1}dp_m/dm \approx 1-4\eta \int_m^{\infty} p_{m'} dm',
$$
or
\begin{equation} \label{continuous approx}
\frac{d \ln{p_m}}{dm} \approx -4\eta \int_m^{\infty} p_{m'} dm'.   
\end{equation}
Guessing the solution of this integro-differential in the form of a power law, we find:
\begin{equation} \label{power law}
    p_m=\frac{1}{2(2\phi-1)m^2}.
\end{equation}

\begin{figure}
    \centering
\includegraphics{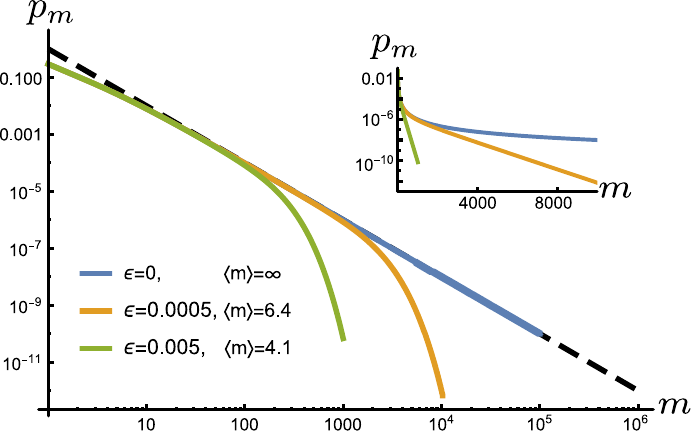}
    \caption{Probability distribution $p_m$ obtained using Eqs.~\eqref{map1}-\eqref{map3} for $\phi=3/4$ and for different values of $\epsilon = p_0-p_c$ and, accordingly, for different $\langle m\rangle$, as indicated. The dashed line is the prediction of Eq.~\eqref{power law}. Inset shows the same data on log scale, demonstrating exponential  distribution tails (cf. Eq.~\eqref{exp tail1}). Note that in the intermediate range where the power law scaling holds (Eq.~\eqref{power law}), the probabilities $p_m$ are independent of the average amount of money per player, $\langle m\rangle$.} 
    \label{fig:power law}
\end{figure}

\begin{figure}
    \centering
\includegraphics[scale=.75]{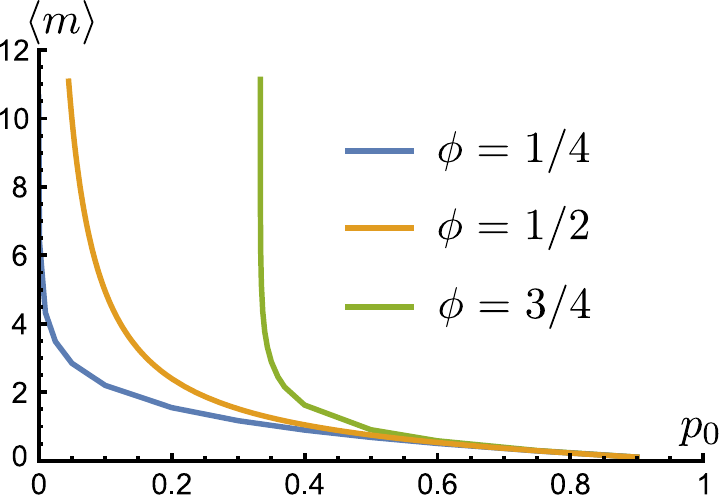}
    \caption{Average amount of money  $\langle m \rangle$ per participant plotted as a function of the probability $p_0$ to have zero money for $\phi=1/4$ (blue), $\phi=1/2$ (orange) and $\phi=3/4$ (green). Note that, in the last case, $\langle m \rangle$ diverges as $p_0$ approaches a finite value of $p_c=1-1/2\phi =1/3$.}
    \label{fig:m vs p0}
\end{figure}

For a nonzero (but sufficiently small) $\epsilon$, therefore, the shape of the curve $p_m$ vs.~$m$ includes an exponential tail at $m\rightarrow\infty$ and an intermediate power-law regime (Fig.~\ref{fig:power law}). When $p_0=p_c$ ($\epsilon = 0$) the exponential tail disappears, and the power law of Eq.~\eqref{power law} holds even in the limit $m\rightarrow\infty$. Such a power law for the tail of the wealth distribution is an example of Pareto's law \cite{pareto1964cours, doi:10.1073/pnas.79.10.3380}, which has been confirmed empirically in various settings \cite{DRAGULESCU2001213, chakrabarti2013econophysics}. In this case, the first moment of the distribution $p_m$ (as well as its higher moments) diverges, since $\int^{\infty}_{0} m p_m dm = \infty$. More generally, as shown in Fig.~\ref{fig:m vs p0}, $\langle m \rangle$ increases and diverges as $p_0$ approaches the critical value $p_c$ from Eq.~\eqref{normCond}. In contrast, for $\phi \le 1/2$, $\langle m \rangle$ diverges as $p_0$ approaches zero (Fig.~\ref{fig:m vs p0}).  

Remarkably, the intermediate power law predicted by Eq.~\eqref{power law} is independent of the value of $p_0$, or, equivalently, of the average amount of money $\langle m \rangle$ per player, and is only a function of the "inequality" parameter $\phi$. This prediction is confirmed by the numerical results for $p_m$ shown in Fig.~\ref{fig:power law}.

Recall that our unit of money here is the amount of money exchanged in each transaction. It is instructive to rewrite our result using more natural units, where the amount exchanged is $\Delta \mu$ and a player's wealth is $\mu = m \times \Delta \mu$ (see Section \ref{SecIII}). In this case we find, for the probability density $p_{\mu}$ of $\mu$ (treating, again, $\mu$ as continuous),
\begin{equation} \label{Pareto}
    p_{\mu}(\mu)\approx p_m \frac{dm}{d\mu}\bigg|_{m=\mu/\Delta\mu} =\frac{\Delta \mu}{2(2\phi-1)\mu^2}.
\end{equation}
Unlike Eq.~\eqref{Boltzmann3}, this result depends, explicitly, on the amount of money $\Delta \mu$ exchanged in each transaction, and the average properties, such as the average wealth $\langle \mu \rangle$, are insufficient for determining the wealth distribution in this case. For example, if we ask how many individuals have wealth below some predefined value $\mu^{*}$, the answer will depend, explicitly, on $\Delta \mu$, not just on $\langle \mu \rangle$. The proportionality of the power-law probability distribution of Eq.~\eqref{Pareto} to $\Delta \mu$ also follows from dimensional arguments: Since $\Delta \mu$ is the only characteristic money scale of the power-law distribution (for which $\langle \mu \rangle = \infty$), the only way the power-law dependence of the probability density on $\mu$, $p_{\mu}(\mu) \propto \mu^{-2}$, can be reconciled with its units of inverse money is that it is also proportional to $\Delta \mu$.

\section{The $\phi>1/2$, finite $N$ case: breakdown of mean-field theory, bimodality of the wealth distribution and the winner-takes-all scenario}\label{SecVII}

\begin{figure}
    \centering
\includegraphics{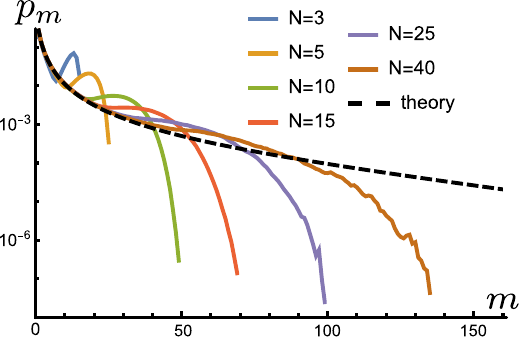}
    \caption{Probability distribution $p_m$ of the money belonging to a player for a game with $N$ players, with $\phi=2/3$ and $\langle m\rangle=5$. Simulations of the game with different values of $N$ are compared with the $N\rightarrow\infty$ result predicted by Eqs.~\eqref{map1}, \eqref{map2}, and \eqref{map3}.}
    \label{fig:bimodal}
\end{figure}

For $\phi> 1/2$, the difference between the case of a finite number of players $N$ and the $N\rightarrow\infty$ limit is not merely quantitative: unlike the $\phi\le 1/2$ case, where the shape of the dependence of $p_m$ on $m$ remains the same regardless of $N$ (cf. Fig.~\ref{fig:mean field}), the distribution $p_m$ may become bimodal (see Fig.~\ref{fig:bimodal}). Importantly, the location of the rightmost peak of the distribution is comparable to $M=N\langle m \rangle$, the total amount of money in the system. Thus, the assumption that $p_m$ is vanishingly small for $m \approx M$ is violated. As $N$ increases, the right distribution peak is shifted to the right, and the probability distribution $p_m$ converges towards the result predicted by mean-field theory (Fig.~\ref{fig:bimodal}).  

\begin{figure}
    \centering
\includegraphics[width=\textwidth]{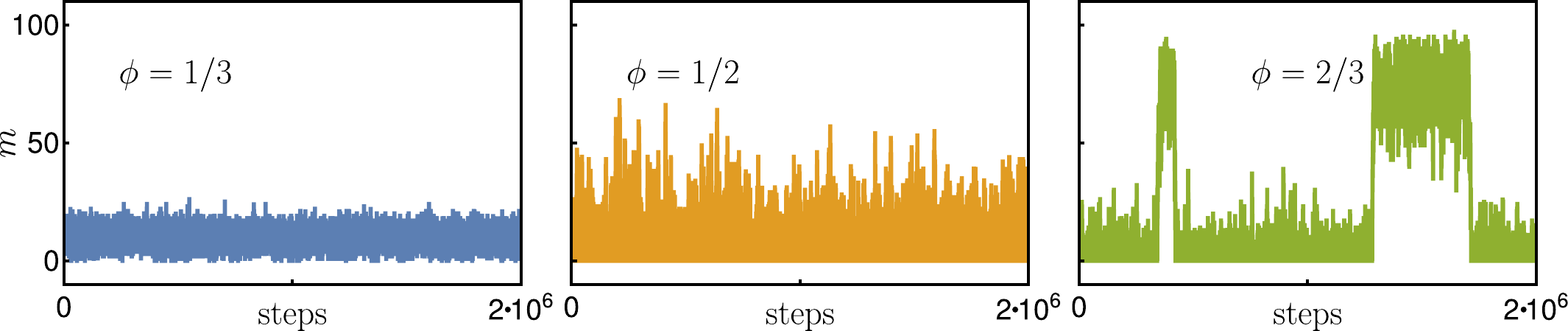}
    \caption{Tracking the money belonging to a player for a segment of time for the three game types, all with the same parameters, $N=10$ and $\langle m \rangle=10$. Note the two distinct states dividing the rich and poor in the $\phi=2/3$ game on the right.}
    \label{fig:bistable dynamics}
\end{figure}

In what follows, it will be shown that this second peak accounts for the possibility that a single player accumulates nearly all the money in the game, thus necessarily forcing the rest of the players into a ``poor'' state. Indeed,   
once a player happens to accumulate more than half of the total money, $M/2$, the rest of the players have less than $M/2$. Under the rules of the game with $\phi>1/2$, the ``rich'' player will be more likely to gain rather than lose money in subsequent exchanges. This leads to a ``winner-takes-all'' scenario until an improbable sequence of money exchanges causes the lucky winner to lose enough money that reverts this player to poverty.  For a sufficiently long game, this will happen, occasionally, to every player. This is illustrated in Fig.~\ref{fig:bistable dynamics}, where the amount of money belonging to a selected player exhibits bistable behavior for $\phi>1/2$ (right panel in Fig.~\ref{fig:bistable dynamics}). In contrast, for $\phi\le 1/2$, no extreme values of $m$ close to $M(=100)$ occur with significant probability (left and middle panels in Fig.~\ref{fig:bistable dynamics}). 

\begin{figure}
    \centering
\includegraphics{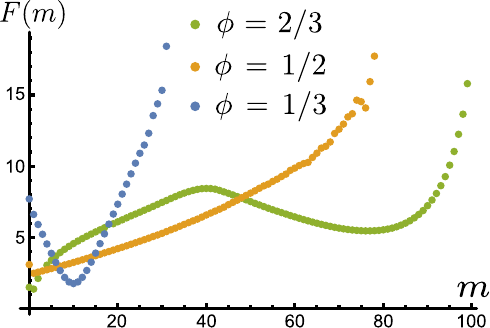}
    \caption{``Free energy'' defined as $-\ln{p_m}$, for the three regimes of the game corresponding to three values of $\phi$, as indicated. In all cases, $N=10$ and $\langle m\rangle=10$. Note the second minimum for the free energy observed for $\phi=2/3$. This minimum corresponds to a value of $m$ that is comparable to the total amount of money in the system, $M=100$ }
    \label{fig:free energy}
\end{figure}

The above points become even more clear when the probability distribution $p_m$ is considered. Equivalently, Fig.~\ref{fig:free energy} shows the ``free energy'', defined as 
\begin{equation} \label{free energy}
F(m)=-\ln p_m     
\end{equation}
as a function $m$. Introduction of the free energy offers a convenient physical picture of the evolution of money $m(t)$ as motion on a free energy surface $F(m)$, which will be explored further below. 

For $\phi\le1/2$, the free energy has a single minimum. For $\phi=1/2$ (the Boltzmann case) this minimum is located at $m=1$, with the probability $p_m$ decaying exponentially (and thus $F(m)$ increasing linearly) as $m$ increases. Thus, the probability of large values of $m$ is exponentially small. For $\phi <1/2$ the single minimum is, roughly, comparable with $\langle m \rangle$. Again, the probability $p_m$ decays quickly to the right of this maximum, making states with $m\gg \langle m \rangle$ improbable. But the case $\phi>1/2$ is qualitatively different, with a second free energy minimum (i.e., probability maximum) corresponding to the player being in a ``rich'' state. 

 Since the total amount of the game's money, $\sum_{i=1}^{N} m_i = M$, is conserved, the existence of a super-rich player (say $j$) that accumulates more than half of the total money ($m_j> M/2$) will subjugate the rest of the players to the poor state with $m_i<M/2, i\ne j$. This means that the trajectories of two different players, $m_i(t)$ and $m_j(t)$, must be coupled. The lack of statistical independence of $m_i(t)$ and $m_j(t)$ implies violation of the assumptions of the mean-field theory -- and, indeed, mean-field theory predicts a monotonic rather than bimodal distribution $p_m$.

\begin{figure}
    \centering
\includegraphics[scale=0.55]{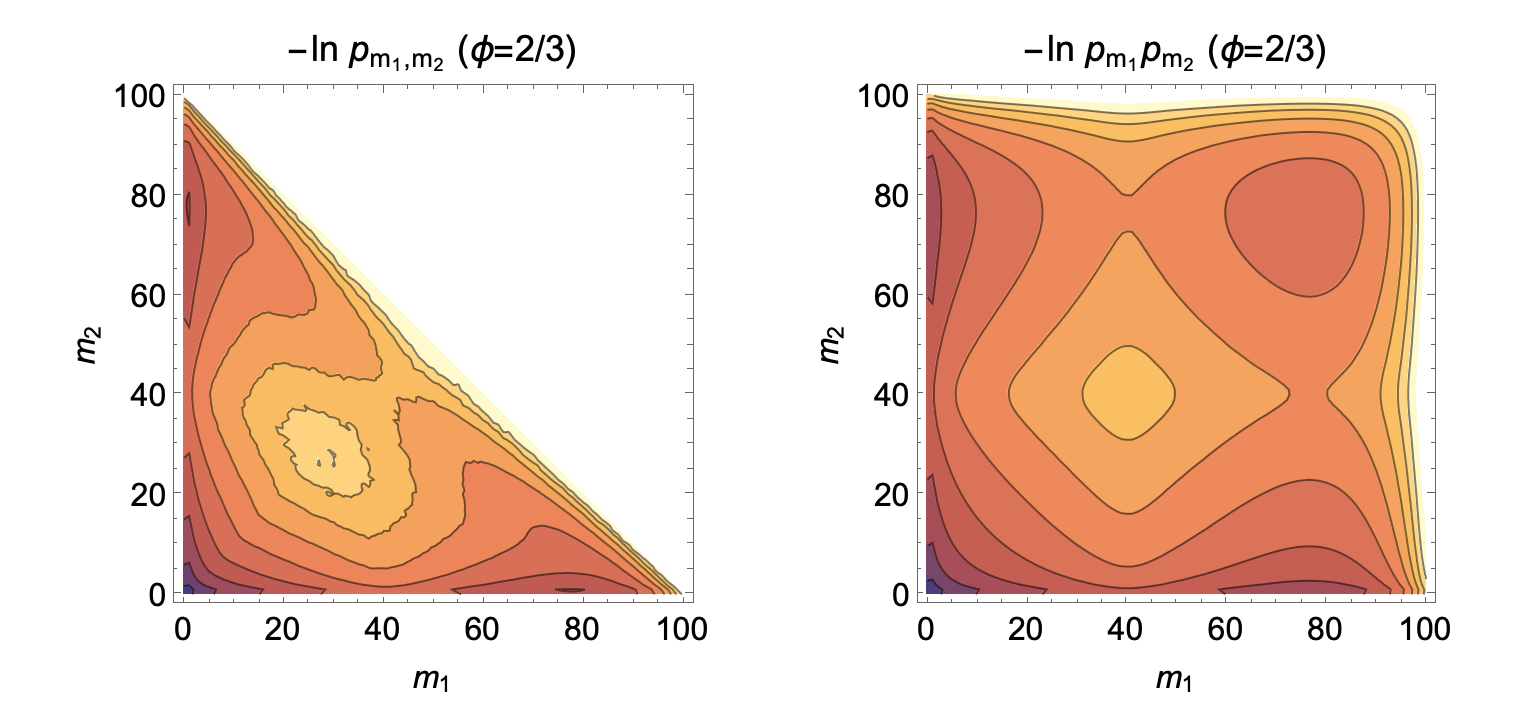}
\includegraphics[scale=0.55]{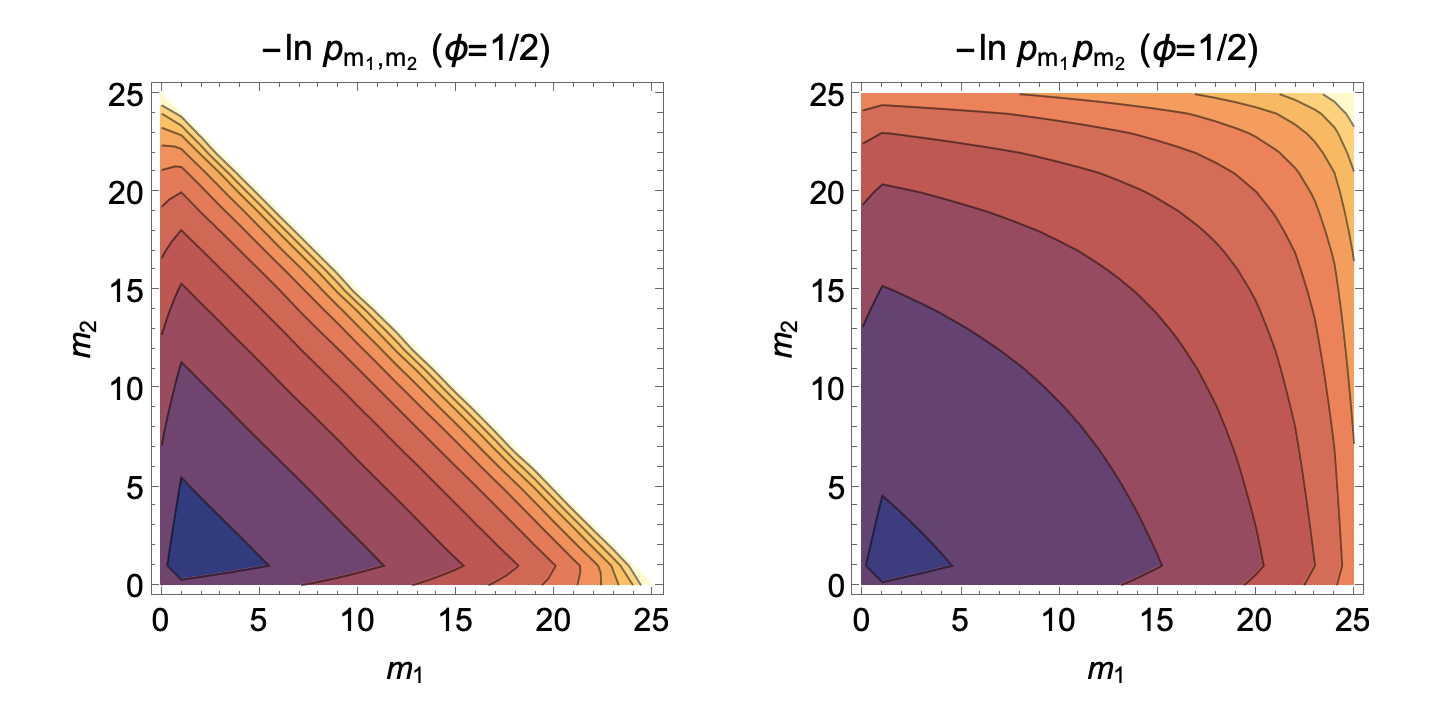}
\caption{Left: contour plots of the joint distribution $p_{m_1,m_2}$ of the money belonging to a pair of players arbitrarily labeled $i=1$ and $j=2$. Right: contour plots of the product of single-player distributions $p_{m_1} p_{m_2}$. If $m_1$ and $m_2$ are statistically independent, then $p_{m_1,m_2}= p_{m_1} p_{m_2}$. Darker color (more violet) correspond to larger values of the probability. As indicated, the top two plots correspond to $\phi=2/3$ and the bottom two plots to $\phi=1/2$. For $\phi=2/3$, the local maximum of the probability product $p_{m_1} p_{m_2}$ situated around $(m_1, m_2)= (80,80)$  represents an impossible state, as the total amount of money in this game is constrained to 100 units. This indicates that $m_1$ and $m_2$ are not statistically independent. Such a maximum is notably absent in the $\phi=1/2$ game. Parameters are $N=10$, $\langle m \rangle=10$ for the $\phi=2/3$ game and $N=5$, $\langle m \rangle=5$ for the $\phi=1/2$ game.}
    \label{fig:distributions2D}
\end{figure}

The coupling between $m_i(t)$ and $m_j(t)$ can be further examined by  comparing their joint distribution $p_{m_i,m_j}$ with the product of single-player distributions $p_{m_i} p_{m_j}$. Such a comparison is given in Fig.~\ref{fig:distributions2D}.   

If the moneys belonging to players 1 and 2 are statistically uncoupled, then their joint distribution is the product of the single-player distributions. The simulation results, however, show this is not the case, with $p_{m_1,m_2}$ qualitatively different from $p_{m_1} p_{m_2}$ for $\phi>1/2$ (Fig.~\ref{fig:distributions2D}). In particular, the product $p_{m_1} p_{m_2}$ has an impossible local minimum at $m_1=m_2>M/2$, where the money belonging to the two players exceeds the total amount of money $M$ in the game. 

To examine, more systematically, how the coupling among players depends on the total number of players, the value of $\phi$, and on the amount of money in the game, we quantify the coupling between players $i$ and $j$ ($i\ne j$) using the mutual information defined as \cite{RN2208}:

\begin{equation} 
I_{ij}=\sum_{m_i=0}^{M}\sum_{m_j=0}^{M}p_{m_i,m_j} \ln\frac{p_{m_i,m_j}}{p_{m_i} p_{m_j}}.
\end{equation}
\\
If $m_i$ and $m_j$ are statistically independent then $p_{m_i,m_j}=p_{m_i} p_{m_j}$, and so $I_{ij}$=0;  
non-zero values of mutual information indicate coupling between $m_i$ and $m_j$. Of course, since all players are equivalent, $I_{ij}$ is independent of $i$ and $j$. 

\begin{figure}
    \centering\includegraphics[scale=1]{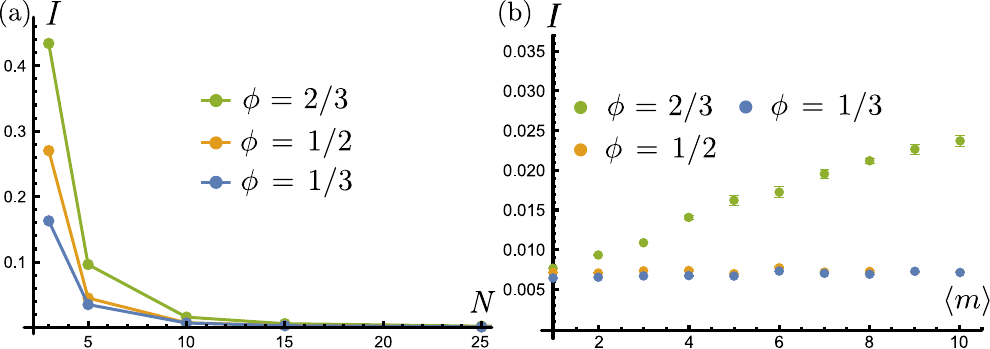}
    \caption{(a) Mutual information between the amounts of money possessed by two players as a function of the number of players \textit{N} in the game for $\phi=1/2$, $\phi=2/3$, and $\phi=1/3$, with $\langle m\rangle=5$. (b) Mutual information between the amounts of money possessed by two players as a function of the average money per player $\langle m\rangle$. The number of players is $N=10$. Increasing money increases the coupling among players in the rich biased game ($\phi=2/3$), but does not have a significant effect on the mutual information for the $\phi=1/2$ and $\phi=1/3$ games.}
    \label{fig:MI}
\end{figure}

As seen from Fig.~\ref{fig:MI}a and, especially, Fig.~\ref{fig:MI}b, the mutual information between pairs of players is significantly greater in the rich-biased games ($\phi=2/3$), as compared to the fair and poor-biased game, given the same number of players. This mutual information further shows significant dependence on the average amount of money accumulated by a player for $\phi>1/2$, but not for $\phi \le 1/2$ (Fig.~\ref{fig:MI}b), in accord with the above qualitative argument suggesting strong coupling in this case. 
In all cases, the mutual information decreases with the number of players $N$ (given a fixed average amount of money per player). This, again, is consistent with the expectation that the mean-field theory prediction should be recovered for $N\rightarrow\infty$.

\section{The finite $N$ case: local equilibrium approximation}\label{SecVIII}

\begin{figure}
    \centering\includegraphics[scale=0.7]{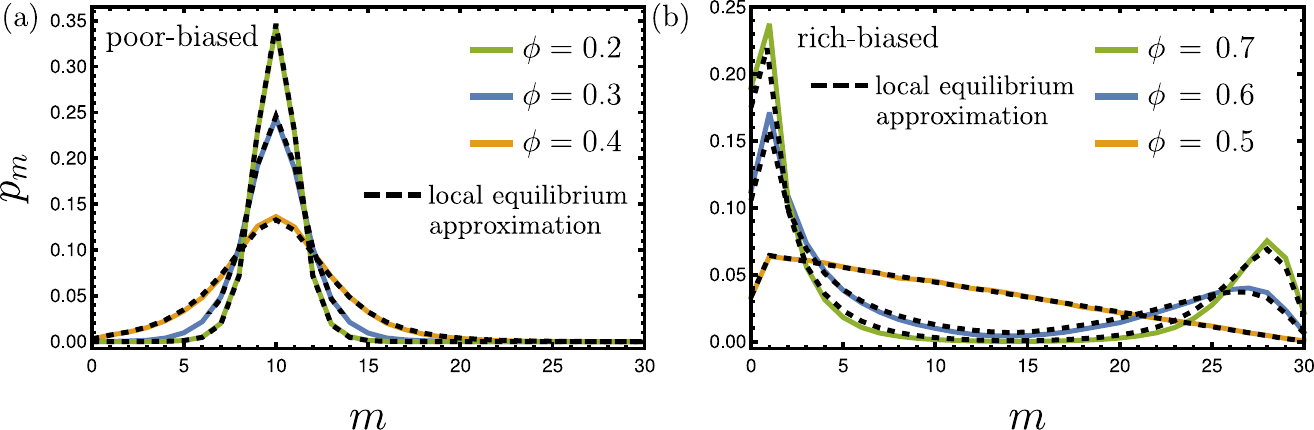}
    \caption{Probability distribution $p_{m}$ of the money belonging to a single player for a poor-biased game (a) and rich-biased game (b) with $N=3$ players and $\langle m \rangle = 10$. Solid lines are obtained from simulation, and dashed lines are given by the local equilibrium approximation.}
    \label{figleq}
\end{figure}

The mean field theory-type approximation described above works in the limit $N\rightarrow\infty$ and cannot explain, even qualitatively, the bistable dynamics and the winner-takes-all scenario occurring in the rich-biased game when the number of players is finite. To account for such finite-size effects, we developed a \emph{local equilibrium} approximation,  which assumes that the dynamics of, for example, player $i$ (with money $m_{i}$) takes place in an equilibrated pool of $N-1$ players with total amount of money $M-m_{i}$. In contrast to the mean field approximation, the local equilibrium approximation explicitly accounts for a finite $M$, and therefore the pool of $N-1$ players does not effectively have infinite money. Similarly to the mean field approximation, the local equilibrium approximation describes the time evolution of the money owned by any given player as a one-dimensional random walk (Fig.~\ref{fig:scheme}), where the effective transition rates are determined by the steady-state probabilities of the $N-1$-player game. An explicit expression for the transition rates and steady-state probabilities in the local equilibrium approximation are given in the Supplementary Material. Comparison of the local equilibrium approximation with simulations shows nearly perfect agreement (Fig.~\ref{figleq}). In particular, the local equilibrium approximation captures the bimodal distribution of money for $\phi>1/2$ (Fig.~\ref{figleq}b).

\section{The $\phi>1/2$, finite $N$ case: non-Markov effects in barrier crossing dynamics}\label{SecIX}

The theory described in Section \ref{SecV} projects $N$-dimensional dynamics onto a single degree of freedom and considers a one-dimensional random walk performed by the money $m(t)$ owned by a single player. Importantly, despite the nonequilibrium character of the $N$-dimensional dynamics (when $\phi \ne 1/2$), the corresponding one-dimensional random walk always obeys detailed balance. The nonequilibrium character of the underlying dynamics is therefore not directly observable in the trajectories $m(t)$, even though it is indirectly reflected in the probability distributions $p_m$ or, equivalently, in the {\em effective} free energies, Fig.~\ref{fig:free energy}. The one-dimensional dynamics $m(t)$ can, in this case, be viewed as one-dimensional random walk/diffusion in the presence of the effective potential $F(m)$. 

The situation is different at finite values of $N$, particularly in the case (considered in Section \ref{SecVII}) where the potential $F(m)$ is bistable. Qualitatively, we can still think of the dynamics along $m$ as a one-dimensional random walk governed by the bistable potential $F(m)$. But because of the coupling between different random walkers, as discussed in Section \ref{SecVII}, each individual random walk no longer has the Markov property.

To illustrate this non-Markov behavior, here we analyze the transition paths between the poor and rich states. A transition path \cite{RN1089,RN2029} is defined as a segment of a single-player trajectory $m(t)$ that enters a ``transition region'' between the poor and rich states, i.e. a segment $(m_P,m_R)$, through its ``poor'' boundary $m_P$ and stays continuously inside this region until exiting through the ``rich'' boundary $m_R$. Analyzing transition paths informs one, for example, whether the "mechanism" of getting rich is the same as that of getting poor: if this is the case, then the ensemble of rich-to-poor transition paths is statistically the same as the ensemble of time-reversed poor to rich transition paths. As a result, then, the mean transition path time (i.e. the mean temporal duration of the transition path) would be the same for the forward and backward transition path, which is often taken as one of the fingerprints of time reversal symmetry.

For a Markovian one-dimensional random walk $m(t)$ this forward-backward symmetry always holds because, as noted above, such a random walk always satisfies detailed balance and thus has time-reversal symmetry \cite{kolmogoroff1936theorie}. But for a non-Markovian random walk, such symmetry may be violated \cite{SashaNonEq}. In all the simulations reported here, we have not be able to find any difference between the mean forward and backward transition path times, to within simulation errors. This, of course, should not be taken as a {\em proof} of the time reversal symmetry of the random walk.  

\begin{figure}
    \centering
\includegraphics[scale=1]{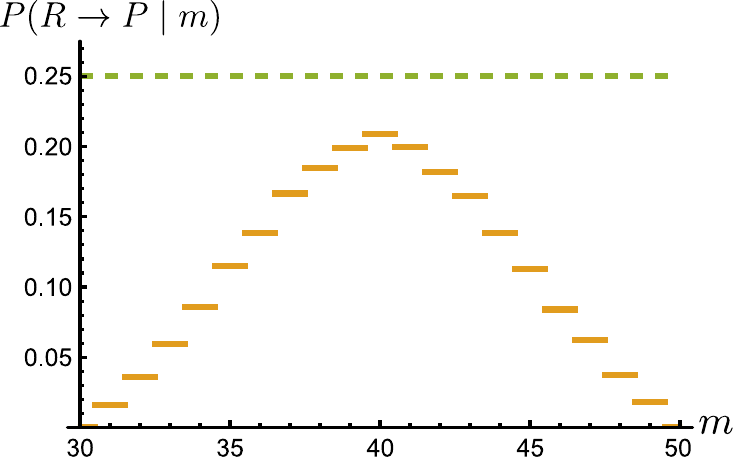}
    \caption{Probability (orange lines) that $m$ belongs to a transition path from the poor ($m_P=30$) to the rich ($m_R=50$) states shows that the dynamics of barrier crossings is non-Markov. For Markovian dynamics, the maximum would be $1/4$ \cite{RN1825}, indicated by the green dashed line.} 
    \label{fig:pTP}
\end{figure}

Another property of the transition path ensemble can be used as a test of Markovianity of the process $m(t)$: consider the probability $P(R\rightarrow P\mid m)$ that a point $m$, $m_P<m<m_R$, belongs to a transition path from the rich to the poor state (as opposed to a trajectory that enters and exits the transition region through the same boundary): for Markovian dynamics this probability attains a maximum value of $1/2 \times 1/2=1/4$ corresponding to a committor-one-half point $m$ where the trajectory starting from $m$ is equally likely to reach either transition-region boundary \footnote{Strictly speaking, this is exactly true only when $m$ is a continuous variable}. For non-Markov dynamics we expect \cite{RN1825}  $\max_{m_P<m<m_R} P(R\rightarrow P\mid m)<1/4$, and, indeed, this is what we observe in Fig.~\ref{fig:pTP} which shows the non-Markov character of the bistable dynamics $m(t)$ observed for $\phi>1/2$. 

\section{Equilibration times}\label{SecX}
Finally, another interesting dynamical property of our model is the time it takes for the steady-state distribution to set in, particularly in the $N\rightarrow \infty$ limit. We set the time between the money exchanges carried out by a single player as the unit of time; this choice is physically sensible, as a "microscopic" time unit is independent of the ensemble size. With this choice, time effectively corresponds to the number of steps performed by the random walker shown in Fig.~\ref{fig:scheme}. 

Consider first the specific scenario where each player starts with $\langle m \rangle$ money units: the initial distribution (measured across the ensemble of players) is infinitely narrow. As this distribution spreads toward the broader equilibrium distribution, each player will, typically, have to travel a distance along the $m$ coordinate that is comparable to the standard deviation of the equilibrium distribution, 
\begin{equation}
    \Delta m = \sqrt{\langle m^2 \rangle -\langle m \rangle^2}. 
\end{equation}
Given the diffusive dynamics along $m$, and assuming a constant diffusivity $D$, the time to attain the equilibrium distribution should be comparable to
\begin{equation}
   \tau \propto \Delta m^2/D. 
   \label{tau1}
\end{equation}
Although this estimate ignores the possible dependence of the diffusivity on $m$, as well as the fact that diffusion is not free but biased, it highlights an important observation that is insensitive to the above assumptions: if $\Delta m$ diverges, then some of the players will have to travel infinitely far from their starting point $m=\langle m \rangle$, which will take infinite time, and so we expect the equilibration time $\tau$ to diverge \footnote{Note that the diffusion coefficient remains finite in this case, given the discrete time unit adopted here}. 

In particular, for the Boltzmann distribution, Eqs.~\eqref{Boltzmann1} and \eqref{Boltzmann2}, the standard deviation is given by
\begin{equation}
    \Delta m=1/\beta = \langle m \rangle.
    \label{dm1}
\end{equation}
Therefore, we expect the relaxation time to diverge in the limit $\langle m \rangle \rightarrow \infty$.  Similarly, the power law distribution of Eq.~\eqref{power law} has infinite variance, and thus the relaxation time should become infinite for $\langle m \rangle \rightarrow \infty$ for any value of the unfairness parameter $\phi$ that exceeds 1/2. The case $\phi<1/2$ is different: using Eqs.~\eqref{map1}-\eqref{map3}, it can be shown that the equilibrium distribution (cf. Fig.~\ref{fig:mean field}) has a finite width. Thus, we expect the equilibration time to be finite even in the $\langle m \rangle \rightarrow \infty$ limit. 

To study the relaxation timescales of the money game more quantitatively, let us first consider the relaxation dynamics for the single-player variable $m$, as revealed by its equilibrium autocorrelation function
\begin{equation} \label{m autocorr}
    \langle m(t) m(0) \rangle = \sum_{m,m'} m' P(m',t\mid m,0) m.
\end{equation}
Here $P(m',t\mid m,0)$ is the propagator, i.e., the joint probability to find the player with $m'$ money units at (discrete) time $t$ given that this player had $m$ money units initially. In the matrix form, the one-step propagator is given by $P(m',1\mid m,0)=\textbf{T}_{m',m}$, where

\begin{equation} \label{transition matrix}
    \textbf{T}=\begin{pmatrix}
        p(0|0) & p(0|1)& 0 & \cdots & \cdots \\
        p(1|0) & 0 & p(1|2) &0 &\cdots  \\
        0 & p(2|1) & 0 & p(2|3) &\cdots \\
        \cdots & \cdots & p(3|2) & 0 & \cdots \\
        \vdots & \vdots & \vdots & \vdots & \ddots
    \end{pmatrix},
\end{equation}
is the matrix of transition probabilities corresponding to the kinetic scheme of Fig.~\ref{fig:scheme}. Similarly, we have
$$
P(m',t\mid m,0)=(\textbf{T}^t)_{m',m}.
$$
Using Eq.~\eqref{m autocorr} now, we conclude that the autocorrelation function has a spectral expansion of the general form
\begin{equation} \label{spectral expansion}
\langle m(t) m(0) \rangle = a_0+a_1 \lambda_1^t+a_2 \lambda_2^t + \dots,
\end{equation}
where $\lambda_0=1$, $\lambda_1$, $\lambda_2$, ... are the positive eigenvalues of $\textbf{T}$ arranged in descending order. Unless the coefficient $a_1$ happens to be identically equal to zero, the long-time decay of the correlation function is dominated by $\lambda_1$, and thus we take  
\begin{equation} \label{relaxation time}
    \tau=-\frac{1}{\log \lambda_1}
\end{equation}
to be the characteristic relaxation time of $m$. 

To study global relaxation dynamics of the entire system, one needs to introduce a global order parameter $R$ that describes the system's dynamics. If $R$ is a linear combination of single-player variables $m_i$'s or, more generally, some function of the form \footnote{For example, one could use the variance $N^{-1}\sum_i (m_i-\langle m \rangle)^2 = \sum_i m_i^2 -\langle m \rangle^2$ as one such global parameter}
$$
    R=\sum_i \alpha_i f(m_i),
$$
then, taking into account the statistical independence of $m_i$ and $m_j$ for $i\ne j$, it is easy to see that 
$$
    \langle R(t) R(0) \rangle = \sum_i \alpha_i^2 \langle f[m(t)] f[m(0)] \rangle,
$$
and thus the autocorrelation function of $R$ has a spectral expansion of the same form as Eq.~\eqref{spectral expansion}. Therefore, we take Eq.~\eqref{relaxation time} as a measure of both projected and global relaxation dynamics.

\begin{figure}
    \centering
\includegraphics[scale=1]{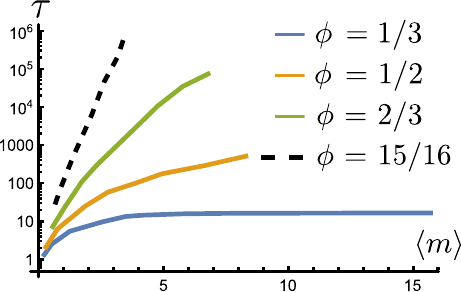}
    \caption{The relaxation time, Eq.~\eqref{relaxation time}, grows indefinitely  as a function of $\langle m \rangle$ for $\phi \ge 1/2$. For $\phi<1/2$ the relaxation time approaches a plateau.}
    \label{fig:relaxation times}
\end{figure}

Figure \ref{fig:relaxation times} shows the relaxation times given by Eq.~\eqref{relaxation time} as a function of the average amount of money per player. Consistent with the arguments above, this time grows indefinitely as $\langle m \rangle \rightarrow \infty$ for $\phi \ge 1/2$ but approaches a plateau for $\phi < 1/2$.  

\section{Concluding remarks}\label{SecXI}

In the money exchange model studied here, the Boltzmann distribution could be derived directly as the steady-state distribution corresponding to the kinetic laws governing the system in the thermodynamic limit ($N=\infty$). It is instructive to consider how the same result can be obtained using conventional statistical mechanics arguments. As the microscopic kinetics of the model is a random walk on a connected graph that is not bipartite, the system is ergodic \cite{Norris1997}. In the case of a ``fair'' game ($\phi = 1/2$), for most connected pairs of microscopic states (namely, those with nonzero money) (cf. Fig.~2) detailed balance leads to equally populated states. Therefore, the principle of  equiprobability of the microstates holds. We can think of the microscopic description of the game as leading to the microcanonical ensemble. Now, if we view an individual player as a small subsystem exchanging money/energy with the large ``reservoir'' formed by the remaining $N-1$ players, then the Boltzmann distribution of this player's money can be deduced using the usual arguments of statistical mechanics \cite{Khinchin_Gamow_2014,LandafshitzStat,Kardar_2020}. 

For $\phi \ne 1/2$, time-reversal symmetry and detailed balance are broken, and the microstates of the system are no longer equally populated -- a microcanonical ensemble description is no longer applicable. As a result, the distribution of an individual player's wealth is no longer exponential. The poor-biased ($\phi < 1/2$) and rich-biased ($\phi > 1/2$) cases turn out to be qualitatively different. The former leads to a bell-shaped distribution of money centered around the mean, which is attained over a finite equilibration time, even when $N=\infty$. The latter leads, in the $N\rightarrow\infty$ limit, to a
broad distribution with a Pareto-type power-law intermediate regime and an exponential tail, such that the first moment of the distribution $\langle m \rangle$ remains finite. As $\langle m \rangle$ increases, the range of values of $m$ for which the power law holds becomes broader, with the exponential tail shifting toward large values of $m$. Remarkably, the probability distribution of money in the power-law regime is independent of the total average wealth $\langle m \rangle$ of a player, and thus of the total amount of money in the game -- that is, the wealth of those ``moderately successful players'' is unaffected by the total wealth. At the same time, the fraction of players in the ``ground state'' with exactly zero money remains finite regardless of the value of $\langle m \rangle$,  and it cannot be lower than a certain critical value $p_c$.  In the limit $\langle m \rangle=\infty$ the power law holds even for $m\rightarrow \infty$, and the distribution's first moment diverges accordingly. The equilibration time also diverges in this case.  

For a finite number of players $N$, the rich-biased game has an interesting regime with bistable dynamics, with each player hopping between a ``super-rich'' state in which they accumulate nearly the entire money pool and a ``poor'' state. This bistability can be captured within the local equilibrium approximation discussed in Section \ref{SecVIII}.  

Finally, we note that the time evolution of the money belonging to a single player (or, analogously, the energy of a single particle exchanging energy with other particles), can be viewed as a projection of a highly multidimensional process (in full state space, as in Fig.~\ref{fig:full kinetics}) onto a single degree of freedom. Such projected dynamics is generally expected to be a non-Markov process \cite{Zwanzig}. But for our system, non-Markov effects are only significant for intermediate values of $N$, particularly in the non-equilibrium case of a rich-biased game, where the winner-takes-all scenario leads to coupling between players. Indeed, such coupling only exists at finite values of $N$ where the total amount of money in the game is finite (Section \ref{SecVII}); on the other hand, for $N=2$ the dynamics of the money belonging to each player is strictly Markovian, as the conservation of money, $m_1+m_2=M$, necessitates that both $m_1$ and $m_2$ are one-dimensional, Markovian random walks (see the Supplementary Material).   

\section*{Acknowledgement} Discussions with Cai Dieball, Irene Gamba, Alja\v{z} Godec, Hagen Hofmann, Matthias Kr\"uger, Peter Sollich, and John Stanton are gratefully acknowledged. This work was supported by the Robert A. Welch Foundation (Grant No. F- 1514 to DEM), the National Science Foundation (Grant No. CHE 1955552 to DEM), and the Alexander von Humboldt Foundation.







\bibliography{mybibliography.bib}

\end{document}


\title{Supplementary Material for: Nonequilibrium statistical mechanics of money/energy exchange models}

\author{Maggie Miao}
\affiliation{Department of Chemistry and Oden
Institute for Computational Engineering and Sciences, The University of Texas at Austin, Austin, Texas 78712, United States}

\author{Kristian Blom}
\affiliation{Mathematical bioPhysics group, Max Planck Institute for Multidisciplinary Sciences, G\"{o}ttingen 37077, Germany}

\author{Dmitrii E. Makarov}
\email{makarov@cm.utexas.edu}
\affiliation{Department of Chemistry and Oden
Institute for Computational Engineering and Sciences, The University of Texas at Austin, Austin, Texas 78712, United States}
\date{\today}

\maketitle
\noindent Here, we calculate the steady-state probability $p_{m}^{N,M}$ for the money game (i.e.~the Bennati-Dregulescu-Yakovenko game \cite{dragulescu2000statisticalS, RevModPhys.81.1703S}) with a finite number of players $N$ and a total amount of money $M$. We start with a two player game where we can calculate the steady-state probability $p^{2,M}_{m}$ exactly. Thereafter, we consider the $N$ player game and employ the local equilibrium approximation to obtain closed-form expressions for $p^{N,M}_{m}$. Subsequently, we show how the local equilibrium approximation can be used to obtain an explicit analytical result for $p_{m}^{N,M}$ when $\phi=1/2$, which converges to the Boltzmann distribution. 
\subsection{Two player game}
\begin{figure}
    \centering
    \includegraphics[width=0.6\textwidth]{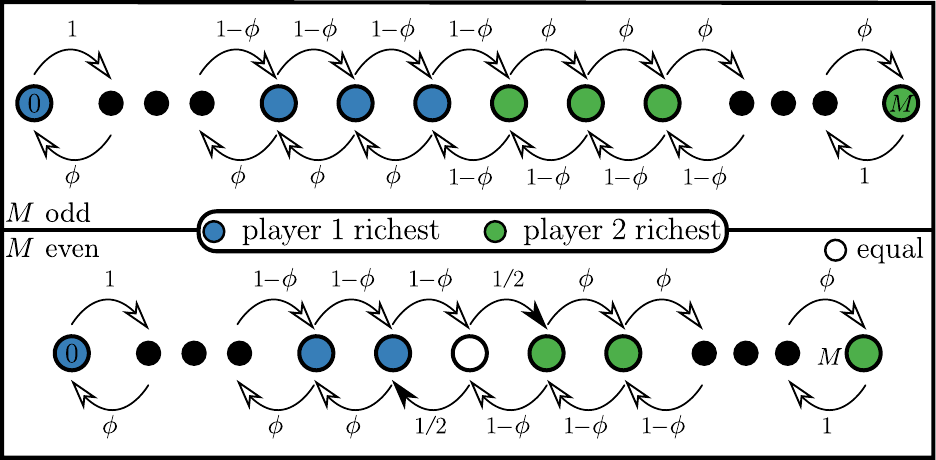}
    \caption{Two player Bennati-Dregulescu-Yakovenko game. In the upper panel, we consider an odd amount of unit money $M$, and in the lower panel an even amount of unit money $M$. Vertices are labeled by the amount of money for player $2$.}
    \label{FigS1}
\end{figure}
\noindent Let us consider the two player game with $N=2$. For a finite amount of money, $M$, the dynamics can be represented by a linear discrete-time Markov chain, as shown in Fig.~\ref{FigS1}. 
Since the Markov chain has a linear structure, the steady-state distribution \emph{for a single player to have $m$ units of money}, denoted as $p^{2,M}_{m}$, can directly be obtained from the detailed-balance relation
\begin{equation}
    p^{2,M}_{m} q^{2,M}_{m\rightarrow m\pm1}=p^{2,M}_{m\pm 1} q^{2,M}_{m\pm1 \rightarrow m},
    \label{DB}
\end{equation}
where $q^{N,M}_{m\rightarrow m\pm1}$ are the conditional probabilities to jump from state $m$ to $m\pm 1$ in the money game with $N$ players and $M$ total amount of money (see Fig.~\ref{FigS1}). Together with the normalization condition, $\sum_{m=0}^{M}p^{2,M}_{m}=1$, this fully specifies the steady-state distribution. For an even amount of money $M$ we obtain from Eq.~\eqref{DB}
\begin{equation}
    p^{2,M}_{m}=\frac{2\phi-1}{4\phi(1-(1/\phi-1)^{M/2})}
    \begin{dcases*}
        1 &, $m=0 \lor m = M$, \\
        2(1/\phi-1)^{M/2-1} &, $m = M/2$, \\
        (1/\phi)(1/\phi-1)^{{\rm min}(m,M-m)-1} &, elsewhere,
    \end{dcases*},
\end{equation}
and for an odd amount of money $M$ we obtain
\begin{equation}
    p^{2,M}_{m}=\frac{2\phi-1}{4\phi-2(1/\phi-1)^{(M-1)/2}}
    \begin{dcases*}
        1 &, $m=0 \lor m = M$, \\
        (1/\phi)(1/\phi-1)^{{\rm min}(m,M-m)-1} &, elsewhere.
    \end{dcases*}
\end{equation}
For $\phi=1/2$ the distribution becomes uniform (with an exception for the tails) 
\begin{equation}
    p^{2,M}_{m}=\frac{1}{2M}
    \begin{dcases*}
        1 &, $m=0 \lor m = M$, \\
        2 &, elsewhere.
    \end{dcases*}
\end{equation}
\subsection{Local equilibrium approximation and the N player game}
\noindent Our next aim is to evaluate the steady-state probability $p^{N,M}_{m}$ for a single player in the money game with $N>2$ players. Unfortunately, due to the nonlinear structure of the Markov chain, in combination with the presence of steady-state probability fluxes, we cannot simply use detailed balance to infer the steady-state probability. To overcome this problem, we will apply the \emph{local equilibrium} approximation, which is based on the assumption that the dynamics between any $N-1$ players is equilibrated before the $N^{\rm th}$ player makes a move. This renders an effective one-dimensional transition landscape for the $N^{\rm th}$ player, where the effective transition rates are determined by averaging over all possible encounters, leading to the following expression 
\begin{equation}
    q^{N,M}_{m\rightarrow m+1}=
    \begin{dcases*}
        \sum_{n=1}^{M}p_{n}^{N-1,M}
        &, $m=0$, \\
        (1/2)p_{1}^{N-1,M-1}+(1-\phi)\sum_{n=2}^{M-1}p_{n}^{N-1,M-1}
        &, $m=1$, \\
        (1/2)p_{m}^{N-1,M-m}+(1-\phi)\sum_{n=m+1}^{M-m}p_{n}^{N-1,M-m}+\phi\sum_{n=1}^{m-1}p_{n}^{N-1,M-m}
        &, $1<m< M/2$, \\
        (1/2)p_{M/2}^{N-1,M/2}+\phi\sum_{n=1}^{M/2-1}p_{n}^{N-1,M/2}
        &, $m=M/2$, \\
        \phi\sum_{n=1}^{M-m}p_{n}^{N-1,M{-}m}
        &, $M/2<m<M$.
    \end{dcases*}.
    \label{weff}
\end{equation}
The backward transitions are simply given by $q^{N,M}_{m\rightarrow m-1}=1-q^{N,M}_{m\rightarrow m+1}$ for $0<m\leq M$. Since the local equilibrium approximation results in an effective one-dimensional transition landscape, the steady-state distribution can directly be determined from the effective transition rates as
\begin{align}
     p^{N,M}_{n}=\left(\delta_{n,0}+\prod_{k=1}^{n}\frac{q^{N,M}_{k-1\rightarrow k}}{q^{N,M}_{k\rightarrow k-1}}\right)/\left(1+\sum_{l=1}^{M}\prod_{k=1}^{l}\frac{q^{N,M}_{k-1\rightarrow k}}{q^{N,M}_{k\rightarrow k-1}}\right).
     \label{p3}
\end{align}
 Note that Eq.~\eqref{weff} explicitly depends on the steady-state probability for the money game with $N-1$ players. Hence, to obtain the steady-state probability of the $N$ player game, we must first obtain the steady-state of the $N-1$ player game (and so forth). For example, for the $3$ player game, we need to insert $p^{2,M}_{n}$ into Eq.~\eqref{weff}. Therefore, the local equilibrium approximation is an iterative procedure to obtain the steady-state probability. 
 
\subsection{N player no-bias game}
\noindent For a biased game with $\phi\neq1/2$ it is rather difficult to obtain closed-form expressions for Eq.~\eqref{p3}, but for $\phi=1/2$ this can be done. For example, for $N=3$ we obtain
\begin{equation}
    p^{3,M}_{m}=\frac{2M}{1+2M^{2}}
    \begin{dcases*}
        1 &, $m=0$, \\
        2(1-m/M) &, $0<m<M$, \\
        1/(2M) &, $m=M$.
    \end{dcases*}
\end{equation}
We can generalize this result to arbitrary $N$. Upon setting $\phi=1/2$ the effective transition rates, given by Eq.~\eqref{weff}, take the simple form
\begin{equation}
    q^{N,M}_{m\rightarrow m+1}=
    \begin{dcases*}
        1-p_{0}^{N-1,M} &, $m=0$, \\
        (1-p_{0}^{N-1,M-m})/2
        &, $1\leq m<M$, \\
        0 &, $m=M$.
    \end{dcases*}
    \label{wnobias}
\end{equation}
Inserting Eq.~\eqref{wnobias} into the general solution Eq.~\eqref{p3}, we obtain the following compact expression
\begin{equation}
     p_{m}^{N,M}=
     \frac{1}{\mathcal{N}_{N,M}}
     \begin{dcases*}
        1 &, $m=0$, \\
        2\frac{1-1/\mathcal{N}_{N-1,M}}{1+1/\mathcal{N}_{N-1,M-m}}\prod_{k=1}^{m-1}\frac{1-1/\mathcal{N}_{N-1,M-k}}{1+1/\mathcal{N}_{N-1,M-k}} 
        &, $1\leq m<M$, \\
        (1-1/\mathcal{N}_{N-1,M})\prod_{k=1}^{M-1}\frac{1-1/1/\mathcal{N}_{N-1,M-k}}{1+1/\mathcal{N}_{N-1,M-k})} &, $m=M$.
    \end{dcases*}
    \label{pgeneral}
\end{equation}
The normalization constant $\mathcal{N}_{N,M}$ is given by the following recursive relation
\begin{equation}
    \mathcal{N}_{N,M}=1+2\sum_{l=1}^{M-1}\frac{1-1/\mathcal{N}_{N-1,M}}{1+1/\mathcal{N}_{N-1,M-l}}\prod_{k=1}^{l-1}\frac{1-1/\mathcal{N}_{N-1,M-k}}{1+1/\mathcal{N}_{N-1,M-k}}+(1-1/\mathcal{N}_{N-1,M})\prod_{k=1}^{M-1}\frac{1-1/\mathcal{N}_{N-1,M-k}}{1+1/\mathcal{N}_{N-1,M-k}},
    \label{N}
\end{equation}
starting with $\mathcal{N}_{2,M}=2M$. Upon iterating Eq.~\eqref{N} we obtain the following results
\begin{equation}
    \mathcal{N}_{3,M}=\frac{1+2M^{2}}{2M}, \  \mathcal{N}_{4,M}=\frac{8M+4M^{3}}{3+6M^{2}}, \  \mathcal{N}_{5,M}=\frac{3+10M^{2}+2M^{4}}{8M+4M^{3}}, \ \mathcal{N}_{6,M}=\frac{46M+40M^{3}+4M^{5}}{15+50M^{2}+10M^{4}},  \ {\rm etc.}
\end{equation}
In the large money limit the normalization constant can be expanded as
\begin{equation}
    \mathcal{N}_{N,M}= \frac{2M}{N-1}+\mathcal{O}\left(\frac{1}{M}\right),
\end{equation}
which after re-inserting into Eq.~\eqref{pgeneral} allows us to determine the steady-state distribution asymptotically
\begin{equation}
    p_{m}^{N,M}= \frac{N-1}{2M}
    \begin{dcases*}
        1 &, $m=0$, \\
        \frac{2(m-M)(N-2(M+1))}{M(N+2(M-m-1))}\frac{(N/2-M)_{m-1}}{(2-N/2-M)_{m-1}} &, $0<m<M$, \\
        \frac{(1+M-N/2)(N/2-M)_{M-1}}{M(2-N/2-M)_{M-1}} &, $m=M$,
    \end{dcases*}
    +\mathcal{O}\left(\frac{1}{M^{3}}\right),
    \label{pnm}
\end{equation}
where $(a)_{n}=a(a+1)\cdot\cdot\cdot(a+n-1)$ denotes the Pochhammer symbol. 
\subsubsection{Convergence to the Boltzmann distribution}
\noindent Equation \eqref{pnm} converges to the Boltzmann distribution (reported in the main text) when we take the large $N$ limit. To see this, let us calculate the moments of Eq.~\eqref{pnm}, which read
\begin{equation}
    \sum_{m=0}^{M}m^{k}p_{m}^{N,M}=\frac{k! M^{k}}{(N)_{k}}+\mathcal{O}(M^{k-1}).
\end{equation}
If we now take the large $N$ limit, we can use $(N)_{k}=N^{k}+\mathcal{O}(N^{k-1})$, to obtain
\begin{equation}
    \sum_{m=0}^{M}m^{k}p_{m}^{N,M}=\frac{k! M^{k}}{N^{k}}+\mathcal{O}((M/N)^{k-1}).
\end{equation}
Here we immediately recognize the moments of the exponential distribution with $M/N \equiv \langle m \rangle$. Therefore, in the limit $N\rightarrow\infty$, while keeping $M/N$ fixed and finite, Eq.~\eqref{pnm} converges to the Boltzmann distribution with temperature $\langle m \rangle$. 
\bibliography{mybibliographyS.bib}